\documentclass[referee]{cjaa}           

\usepackage{graphicx}                   
\input{epsf.sty}                        
\input{psfig.sty}                       

\def\saxj{SAX J1808.4--3658}
\def\igrj{IGR J00291+5934}

\def\Porb{P_{\rm orb}}

\def\T0{T^*_0}
\def\asini{a_1 \sin i}

\begin{document}

   \title{Timing an Accreting Millisecond Pulsar: Measuring the Accretion
Torque in \igrj 
}


   \author{L. Burderi
      \inst{1}\mailto{burderi@mporzio.astro.it}
   \and T. Di Salvo
      \inst{2}
   \and A. Riggio
      \inst{2}
   \and M.T. Menna
      \inst{3}
   \and G. Lavagetto
      \inst{2}
   \and A. Papitto
      \inst{3}
   \and R. Iaria
      \inst{2}
   \and N. R. Robba
      \inst{2}
   \and L. Stella
      \inst{3}
      }
   \offprints{L. Burderi}                   

   \institute{Universit\`a degli Studi di Cagliari, Dipartimento
	      di Fisica, SP Monserrato-Sestu, KM 0.7, 09042 Monserrato, Italy\\
             \email{burderi@mporzio.astro.it}
        \and
             Dipartimento di Scienze Fisiche ed Astronomiche,
	     Universit\`a di Palermo, via Archirafi 36 - 90123 Palermo, Italy\\
        \and I.N.A.F. - Osservatorio Astronomico di Roma, via Frascati 33,
	     00040 Monteporzio Catone (Roma), Italy\\
          }

   \date{Received~~2005 month day; accepted~~2005~~month day}

   \abstract{We present here a timing analysis of the fastest accreting millisecond 
pulsar \igrj\ using RXTE data taken during the outburst of December 2004. 
We corrected the arrival times of all the events for the orbital (Doppler) effects 
and performed a timing analysis of the resulting phase delays. In this way we find 
a clear parabolic trend of the pulse phase delays showing that the
pulsar is spinning up as a consequence of accretion torques during the X-ray
outburst.  The accretion torque gives us for the first time an independent
estimate of the mass accretion rate onto the neutron star, which can be compared
with the observed X-ray luminosity.
We also report a revised value of the spin period of the pulsar.
}

   \authorrunning{L. Burderi et al. }            
   \titlerunning{Timing the Fastest Accreting Millisecond Pulsar }  


   \maketitle
%
%
\section{Introduction}           
\label{sect:intro}


The so-called recycling scenario links two different classes of
astronomical objects, namely the millisecond radio pulsars (usually
found in binary systems) and the Low Mass X-ray Binaries (hereafter LMXBs),
or, at least, a subgroup of them. The leading idea of this scenario
is the recycling process itself, during which an old, weakly
magnetized, slowly spinning neutron star is accelerated by the accretion
of matter and angular momentum from a (Keplerian) accretion disk down
to spin periods in the millisecond range. In this way, at the end of
the accretion phase, the neutron star rotates so fast that it is
resurrected from the radio pulsar graveyard, allowing the radio pulsar
phenomenon to occur again despite the weakness of the magnetic field.

Although this scenario was firstly proposed long time ago (see
e.g. Bhattacharya \& van den Heuvel 1991 for a review), the most embarassing
problem was the absence of coherent pulsations in LMXBs. Only recently, the
long seeked for millisecond coherent oscillations in LMXBs have been found,
thanks to the capabilities (the right combination of high temporal resolution
and large collecting area) of the RXTE satellite. In April 1998, a transient
LMXB, \saxj, was discovered to harbour a millisecond pulsar ($P_{\rm spin}
\simeq 2.5$~ms) in a compact ($P_{\rm orb} \simeq 2$~h) binary system
(Wijnands \& van der Klis 1998; Chakrabarty \& Morgan 1998).  We now know
seven accreting millisecond pulsars (Wijnands 2005; Morgan et al. 2005); all
of them are X-ray transients in very compact systems (orbital period between
40 min and 4 h), the fastest of which ($P_{\rm spin} \simeq 1.7$~ms), \igrj,
has been discovered in December 2004 (Galloway et al.\ 2005, hereafter G05).
Timing techniques applied to data of various accreting millisecond pulsars,
spanning the first few days of their outbursts, allowed an accurate
determination of their main orbital parameters.  However, only a few attemps
have been made to determine the spin period derivative (Chakrabarty et al.\
2003; Galloway et al.\ 2002).

In this paper we apply an accurate timing technique to the fastest currently
known accreting millisecond pulsar, \igrj, in the hope of constraining the
predictions of different torque models with good quality experimental data.
Our results indicate quite clearly that a net spin up occurs during the
December 2004 outburst of \igrj\ (see also Falanga et al.\ 2005) and that the 
derived torque is in good agreement with that expected from matter accreting 
from a Keplerian disk.

\section{The Timing Technique}
\label{sect:data}

In standard timing techniques (see e.g.\ Blandford \& Teukolsky 1976) the
predicted arrival time of a given pulse is computed using a first guess of the
parameters of the system, and the difference between the experimental and
predicted arrival times, namely the residuals, are fitted with a linear
multiple regression of the differential corrections to the parameters.  This
means that the differential correction to orbital parameters, source
position in the sky, spin frequency and its derivative, are computed
simultaneously. This technique has the
obvious advantage to give a self-consistent solution, where all the
correlations in the covariance matrix of the system are fully taken into
account. However, the convergency of the fit is not always guaranteed and --
especially on long temporal baselines -- convergence to secondary minima could
occur.

On the other hand, if the orbital period is much shorter than the timescale on
which the spin period derivative is expected to produce a significant effect,
we can demonstrate that a different timing technique is more effective in
determining the spin period derivative. This technique relies on the fact that
the delays in the arrival times produced by the orbital corrections are
effectively decoupled from those caused by the spin evolution. The technique
proceedes as follows: in order to obtain the emission times, $t_{\rm em}$, the
arrival times of all the events, $t_{\rm arr}$, are firstly reported to
the Solar system barycenter adopting the best estimate of the source
position in the sky, then corrected for the delays
caused by the binary motion using the best estimate of the orbital parameters
through the formula:
\begin{equation}
\label{eq:corr}
t_{\rm em} \simeq t_{\rm arr} - x \sin \left[ \frac{2\pi}{\Porb}
\left(t_{\rm arr} - T^* \right)\right],
\end{equation}
where $x = a \sin i /c$ is the projected semimajor axis in light seconds, and
$T^*$ is the time of ascending node passage at the begining of the
observation. In the following, for simplicity, we use $t$ instead of $t_{\rm
em}$.  The differential of this expression with
respect to the orbital parameters allows to calculate the uncertainties in the
phases, $\sigma_{\phi\,\rm orb}$, caused by the uncertainties in
the estimates of the orbital parameters. 
In a similar way, we have computed the uncertainties in the phase delays,
$\sigma_{\phi\,\rm pos}$, caused by the uncertainties on the estimates of
the source position in the sky. 
In this case we can estimate $\dot \nu$ fitting the measured phase variations, 
while the uncertainties in the adopted orbital parameters and source position 
will result in a ``timing noise'' of amplitude $\sigma_{\phi\,\rm par} =
(\sigma_{\phi\,\rm orb}^2 + \sigma_{\phi\,\rm pos}^2)^{1/2}$ (see Burderi 
et al.\ 2005 for details).

\section{Observations and Data Analysis}
\label{sect:obs}

\igrj\ was observed by RXTE between 2004 December 3 and 21. In this
paper we report on the data between December 7 and 21 taken from a public ToO.
We mainly use data from the PCA for timing anaysis and data from PCA and HEXTE
for spectral analysis.
The arrival times of all the events were converted to barycentric dynamical 
times at the solar system barycenter. The position
adopted for the source was that of the proposed radio counterpart (which is
compatible with that of the proposed optical counterpart, see Rupen et al.\
2004). We corrected the arrival times of all the events for the delays caused by the
binary motion using eq.~(\ref{eq:corr}) with the orbital parameters given in
G05.  

Adopting the uncertainties in the estimates of the orbital parameters given in
G05, the positional uncertainty of $0.04''$ radius reported by Rupen et 
al. (2004) we obtain: 
$\sigma_{\phi\, \rm orb} \la 0.01$, $\sigma_{\phi\,\rm pos} \la 0.006$
where we have
maximized $\sin$ and $\cos$ functions with 1, and used $t-T_0 \la 7$ days.
Therefore, we expect that the uncertainties in the orbital parameters and
source position will cause a ``timing noise'' not greater than $\sigma_{\phi\,
\rm par} \times P_{\rm spin} \sim 0.02$~ms.

To compute phases of good statistical significance we epoch folded each
interval of data in which the pulsation was significantly detected at the spin
period given in G05 with respect to the same reference epoch, $T_0$,
corresponding to the begining of our observations. The fractional part of the
phase was obtained fitting each pulse profile with a sinusoid of fixed
period. To compute the associated errors we combined the statistical errors
derived from the fit, $\sigma_{\phi\,\rm stat}$, with the errors
$\sigma_{\phi\, \rm par}$.

In order to derive the differential correction to the spin frequency,
$\Delta \nu_0$, and its derivative, $\dot \nu_0$, at the time $T_0$ we have to
derive a functional form for the time dependence of the phase delays.  We
started from the following simple assumptions: 
i) the bolometric luminosity $L$ is a good tracer of the mass
accretion rate $\dot M$ {\it via} the relation $L = \zeta
(GM/R) \dot M$, where $\zeta \leq 1$, and $G$, $M$, and $R$ are the
gravitational constant and the neutron star mass and radius, respectively.
ii) The matter accretes through a Keplerian disk truncated at the
magnetospheric radius, $R_{\rm m} \propto \dot M^{-\alpha}$, by its
interaction with the (dipolar) magnetic field of the neutron star.
At $R_{\rm m}$ the matter is forced to corotate with the
magnetic field of the neutron star and is funneled (at least in part) towards
the rotating magnetic poles, thus causing the pulsed emission. For standard
disk accretion $\alpha =2/7$.
Indeed we considered two extreme cases, namely $\alpha = 2/7$ and
$\alpha = 0$, since a location of $R_{\rm m}$ independent of $\dot M$ has
been proposed (see, {\it e.g.}, Rappaport, Fregeau, and Spruit, 2004).
iii) The matter accretes onto the neutron star its
specific Keplerian angular momentum at $R_{\rm m}$, $\ell = (GMR_{\rm
m})^{1/2}$, thus causing a material torque $\tau_{\dot M} = \ell
\times \dot M$. A firm upper limit to this torque is given by the condition
$\tau_{\dot M} \leq \ell_{\rm max} \times \dot M$, with $\ell_{\rm max} =
(GMR_{\rm CO})^{1/2}$, where $R_{\rm CO} = 1.50 \times 10^8 \; m^{1/3}
\nu^{-2/3}$ is the corotation radius (namely the radius at which the Keplerian
frequency equals the spin frequency $\nu$ of the neutron star and beyond which 
accretion is centrifugally inhibited), and $m = M/{\rm M_\odot}$. 
We {\it do not consider} any form of threading of the accretion disk by
the magnetic field of the neutron star (see e.g.\ 
Rappaport, Fregeau, and Spruit 2004 for a description of the magnetic
threading), which implies that the only torque acting during accretion is
$\tau_{\dot M}$.

\begin{table}
\footnotesize
\caption{Orbital and spin parameters of \igrj.}
\label{table:1}
\newcommand{\m}{\hphantom{$-$}}
\newcommand{\cc}[1]{\multicolumn{1}{c}{#1}}
\renewcommand{\tabcolsep}{0.6pc} 
\renewcommand{\arraystretch}{1.2} 
\begin{center}
\begin{tabular}{@{}lll}
\hline
Parameter	 &	G05 	&	This work	\\
\hline
Projected semimajor axis, $\asini$ (lt-ms)                      & $64.993(2)$ &
  --   \\
Orbital period, $\Porb$ (s)                                     & $8844.092(6)$
& --   \\
Epoch of ascending node passage, ${T^*}$ (MJD) & $53345.1619258
(4)$  & --  \\
Eccentricity, $e$                                               & $<2 \times 10^
{-4}$ (3 $\sigma$)  & -- \\
Spin frequency, $\nu_0$ (Hz)                                    & $598.89213064(
1)$ & $598.89213053(2)$ \\
Spin frequency derivative, $\dot \nu_0$ (Hz/s)                    & $< 8 \times
10^{-13} $ (3 $\sigma$) & --  \\
Spin frequency derivative, $\dot \nu_0$ (Hz/s) ($\alpha=0$)  &  --  &
$1.17(0.16)\times 10^{-12} \; \;$    \\
Spin frequency derivative, $\dot \nu_0$ (Hz/s) ($\alpha=2/7$) &  --  &
$1.11(0.16)\times 10^{-12} \; \; $   \\
Spin frequency derivative, $\dot \nu_0$ (Hz/s) ($\dot \nu =$ constant) &  --  &
$0.85(0.11)\times 10^{-12} \; \; $   \\
Epoch of the spin period, ${T_0} $  (MJD)                          &  --     &
53346.184635 \\
\hline
\end{tabular}\\[2pt]
\end{center}
\footnotesize
Errors are given at $1 \sigma$ confidence level.
\end{table}

In these hypothesis the spin frequency derivative is $\dot \nu = \ell \,
\dot M/(2\pi \, I) $,
where $I$ is the moment of inertia of the neutron star and we
have neglected any variation of $I$ caused by accretion.
If $\dot M = \dot M(t)$, we have $\dot \nu(t) = (2\pi
\,I)^{-1}  \ell_0 \dot M_0 (\dot M(t)/\dot M_0)^{1-\alpha/2}$, where 
$\ell_0 = (GMR_{\rm m \, 0})^{1/2}$, and $R_{\rm m \,0}$ and $\dot M_0$ 
are $R_{\rm m}$ and $\dot M$ at $t = T_0$, respectively. 
For the $\alpha = 0$ case we assumed $\ell_0 = \ell_{\rm max}$.

\begin{figure}
   \vspace{2mm}
   \begin{center}
   \hspace{3mm}\psfig{figure=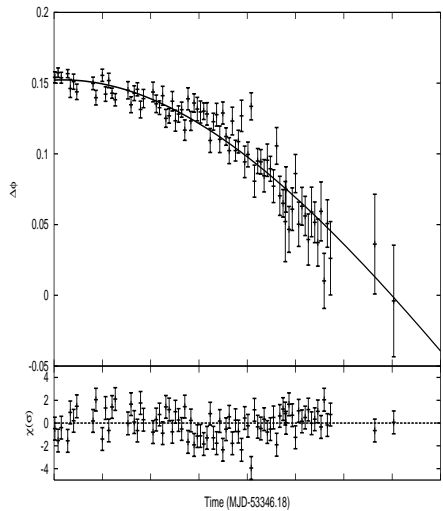,width=60mm,height=70mm,angle=0.0}
   \caption{Pulse phases computed folding at the spin period reported in
Table~1 and plotted versus time together with the best fit curve for
$\alpha = 2/7$ shown as a solid line (upper panel) and residuals in units of
$\sigma$ (lower panel). Note that the linear term is fully compatible
with zero.   }
   \label{fig1}
   \end{center}
\end{figure}

Since we assumed $\dot M(t) \propto L(t)$, to determine the temporal
dependence of $\dot M(t)$ we studied
the energy spectra of the source for each continuous interval of
data combining PCA and HEXTE data. All the spectra are well fitted with a
model consisting of a power law with an exponential cutoff plus thermal
emission from a Keplerian accretion disk modified by photoelectric absorption
and a Gaussian iron line. In order to derive $L(t)$ for each spectrum
we made the simple assumption $L(t) \propto F_{(3-150)}(t)$, which is
the unabsorbed flux in the RXTE PCA plus HEXTE energy band $(3-150$ keV).
A good fit of $F_{(3-150)}(t)$ {\it vs} $t$ between December 7 and 14
($\Delta t_{\rm obs} \sim 7.3$ days) is given by the expression
$F_{(3-150)}(t) = F_{(3-150)} \times [1- (t-T_0)/t_B]$ with
$t_B=8.4 \pm 0.1 $ days, where $F_{(3-150)}$ is the unabsorbed flux
at $t=T_0$.
Therefore we have $\dot \nu(t) = \dot \nu_0 \times
[1-(t-T_0)/t_B]^{1-\alpha/2}$, where the spin frequency
derivative at $t=T_0$ is
$\dot \nu_0 = (2\pi \,I)^{-1} \, \ell_0 \, \dot M_0$.
We have therefore fitted the phase delays with the function:
\begin{equation}
\phi = - \phi_0 - \Delta \nu_0\, (t - T_0) -
\frac{1}{2} \dot \nu_{0} (t - T_0)^2 \times
\left[ 1 - \left(  \frac{2-\alpha}{6} \right) \times
\frac{(t-T_0)}{t_B} \right] .
\label{eq:phi}
\end{equation}
Using the best fit value for $\Delta \nu_0$ we computed the improved spin
frequency estimate and repeated the same procedure described at the begining
of this paragraph, folding at the new estimate of the spin period. The new
phases were fitted with eq.~(\ref{eq:phi}). In this case, $\Delta \nu_0$ was
fully compatible with zero. These phases are plotted versus time in
Figure~\ref{fig1} (upper panel) together with the residuals in units of
$\sigma$ with respect to eq.~(\ref{eq:phi}) (lower panel). The best fit
estimates of $\nu_0$ and $\dot \nu_0$ are reported in Table~1
for three values of $\alpha$, namely $\alpha = 0$ which correspond
to a location of $R_{\rm m}$ independent of the accretion rate
(cfr. the model of Rappaport Fregeau, and Spruit 2004 in which $R_{\rm m}
= R_{\rm CO}$ for any $\dot M$), the standard case $\alpha = 2/7$ which
corresponds to $R_{\rm m}$ proportional to the Alfv\'en radius,
and $\alpha = 2$
which has been given for comparison purposes and corresponds to a parabolic
trend, expected in the case of constant $\dot M$. The statistics
is not good enough to distinguish between these three possibilities.

From the best-fit value of the spin frequency derivative $\dot \nu_0$
we can compute the mass accretion rate at $t = T_0$ through the
formula:
$
\dot M_{-10} = 5.9 \times \dot \nu_{-13}\, I_{45}\,
m^{-2/3} (R_{\rm CO}/R_{\rm m \, 0})^{1/2},
$
where $\dot M_{-10}$ is $\dot M_0$ in units of $10^{-10}\, M_\odot$ yr$^{-1}$,
$\dot \nu_{-13}$ is $\dot \nu_0$ in units of $10^{-13}$ s$^{-2}$, and
$I_{45}$ is $I$ in units of $10^{45}$ g cm$^2$. In the following we will
adopt the FPS equation of state for the neutron star matter for $m = 1.4$
and the spin frequency of \igrj\ which gives $I_{45}= 1.29$ and
$R= 1.14 \times 10^6$ cm (see e.g.\ Cook, Shapiro \& Teukolsky 1994).
In order to compare the experimental estimate of $\dot M_0$ with the
observed X-ray luminosity, we have to derive the bolometric luminosity
$L(t)$ from the observed flux $F_{(3-150)}(t)$.
To this end we consider the spectral shape at $t=T_0$
in more detail.

The power law is the dominant spectral component.
This component presumably originates in an atmosphere of small
optical depth just above each polar cap (see e.g.\ Poutanen \& Gierlinski 2003;
Gierlinski \& Poutanen 2005), thus we neglect, to first order, any effect
of the inclination of the emitting region with respect to the observer.
On the other hand, we observe a single-peaked pulse profile, which means
that we only see the emission from one of the polar caps (e.g.\ Kulkarni \&
Romanova 2005). We have therefore multiplied by 2 the unabsorbed flux of the
power law in order to take into account the emission of the unseen polar cap.
Assuming isotropic emission, we computed 
$L_{{\rm PL}} \simeq 2 F_{{\rm PL} \; (0,\infty)}
\times 4 \pi d^2 = 1.5^{+0.4}_{-0.3} \times 10^{37} \; d_{\rm 5\,kpc}^2$ erg/s,
where $d_{\rm 5\, kpc}$ is the source distance in units of $5$ kpc.
The uncertainty on the luminosity has been evaluated propagating the uncertainties
on the spectral parameters threated as they were independent on each other.

The second component is the thermal emission from a Shakura-Sunyaev accretion
disk. 
The fraction of the total luminosity that is emitted by the disc is given by
the ratio: $0.5 R/R_{\rm m\,0} = 0.34$. In this hypothesis
$L_{{\rm BB\; 0}} = 0.39 / (1-0.39) \times L_{{\rm PL} \; 0}
= 9.6 \times 10^{36} \, d_{\rm 5\,kpc}^2$ erg/s.
The total bolometric luminosity is therefore $L_0 = 2.46^{+0.94}_{-0.29}
\times 10^{37} \, d_{\rm 5\,kpc}^2$ erg/s.
If we compare this luminosity with the mass accretion rate inferred from the
timing analysis, we obtain an estimate of the source distance, which is:
$d = 9.47_{-2.1}^{+1.2}$ kpc.

\section{Conclusions}
\label{sect:conclusion}

We have analysed RXTE data of the fastest known accreting millisecond
pulsar, \igrj, during the period $7 - 14$ December, 2004. We report a revised
estimate of the spin period and the spin period derivative. The source
shows a strong spin-up, which indicates a mass accretion rate of about
$8.5 \times 10^{-9} \, M_\odot$ yr$^{-1}$. 
Comparing the bolometric luminosity of the source as derived from the X-ray spectrum 
with the mass accretion rate of the source as derived from the timing, we find a
good agreement if we place the source at a quite large distance between 7 and
10 kpc. Note that 10 kpc is close to the outer edge of our Galaxy in the direction
of \igrj. Another possibility is that part of the luminosity of the system is
not observed because emitted in other energy bands or because of occultation
effects (which may be favoured if indeed the source is highly inclined).

\begin{acknowledgements}
This work was partially supported by the Ministero della Istruzione,
della Universit\`a e della Ricerca (MIUR).
\end{acknowledgements}

\label{lastpage}

\end{document}